\documentclass[12pt]{article}%
\usepackage{amsmath}
\usepackage{graphicx}
\usepackage{amsfonts}
\usepackage{amssymb}%
\providecommand{\U}[1]{\protect\rule{.1in}{.1in}}
\begin{document}

\title{The Range of \ the Kondo Cloud in Weakly Disordered Hosts}
\author{Gerd Bergmann and Richard S. Thompson\\Department of Physics\\University of Southern California\\Los Angeles, California 90089-0484\\e-mail: bergmann@usc.edu}
\date{\today}
\maketitle

\begin{abstract}
The calculation of the Kondo cloud is extended to disordered hosts. For a
weakly disordered large three-dimensional host the structure of the ground
state is very close to the pure host. However, the range of the disordered
electron basis is much shorter. The extention of the Kondo cloud is
essentially given by $\sqrt{\xi_{K}l}$ where $\xi_{K}$ is the Kondo length in
the pure host and $l$ the mean free path.

 \newpage

\end{abstract}

\section{Introduction}

The properties of magnetic impurities in a metallic host have fascinated
physicists for a long time, particularly after the publication of Kondo's
paper \cite{K8} 50 years ago. In the meantime the field of the "Kondo effect"
has matured (see for example \cite{H20}, \cite{B195}, \cite{W53},
\cite{B187}). One of the open questions is the so-called Kondo cloud. The idea
is to divide the Kondo ground state, the singlet state $\Psi_{SS}$, into two
parts with opposite (net) d-spins. The proponents of the Kondo cloud argue
that in each component there is a cloud of s-electron spins that compensates
the d-spin. Such a cloud has been theoretically derived \cite{H42},
\cite{Y16}, \cite{S84}, \cite{A98}, \cite{A84}, \cite{A81}, \cite{A80},
\cite{W44}, \cite{A82}, \cite{B176}, \cite{S83}, \cite{A83}, \cite{B177},
\cite{B178}, \cite{B181}, \cite{B185}, \cite{D53}, \cite{B182}, \cite{B204},
\cite{B203}, \cite{B201}, \cite{P54}. This predicted Kondo cloud has not yet
been experimentally detected.

If the host is pure (except for the magnetic impurity) then the range of this
cloud is of the order of the Kondo length
\begin{equation}
\xi_{K}=\frac{\hbar v_{F}}{k_{B}T_{K}} \label{K_xi}%
\end{equation}
($k_{B}T_{K}$ =Kondo energy, $v_{F}$ = Fermi velocity of the s-electrons).

One reason why an experimental detection of the Kondo cloud has been so
difficult is the fact that the Kondo cloud is very dilute, about a single spin
distributed over $10^{14}$ host atoms in a pure three-dimensional host with a
$T_{K}$ of a few Kelvin. In this paper we discuss the possibility to reduce
the Kondo length dramatically by using a disordered host with a finite mean
free path $l$. From superconductivity it is well known that a finite mean free
path reduces the BCS coherence length from $\xi_{0}=\hbar v_{F}/\left(
k_{B}T_{c}\right)  $ to a much smaller length of $\xi=\sqrt{\xi_{0}l}$.
Although the underlying physics of the Kondo effect and of superconductivity
are quite different it is worthwhile to investigate the influence of disorder
on the Kondo length and in particular the extension of the Kondo cloud.

The influence of disorder on the Kondo effect is investigated in a number of
papers \cite{S98}, \cite{D58}, \cite{K82}, \cite{K83}. Many focus on the
extreme case of electron localization. In this case the magnetic impurity
interacts only with a relatively small number of localized electrons, which in
a way is similar to the case of small sample size.

In a weakly disordered host in three dimensions the electron density is in
first approximation close to that of a pure host although the disorder causes
some fluctuations. Any such fluctuation at the position of the magnetic
impurity results in a change of the Kondo temperature. Therefore a finite
degree of disorder in the host yields a spread of the Kondo temperatures for
different realizations of the disorder. Such profiles of the Kondo temperature
have been calculated \cite{K82}.

In the present paper we restrict ourselves to small disorder, i.e. $k_{F}%
l>>1$, in three dimensions where the fluctuations are small. We consider a
large host so that the spacing of the energy levels is much smaller than the
Kondo energy. We use the fact that the Kondo cloud is already calculated for
the pure host (see for example \cite{B177}) and show that indeed the Kondo
cloud in a disordered host has a reduced extension of $\sqrt{\xi_{K}l}$. We
derive a relatively simple formula to calculate any polarization in the
disordered host from the polarization in the pure host.

The basis of this paper is the solution of the Friedel-Anderson (FA) impurity
problem in the FAIR approach by our group \cite{B151}, \cite{B152},
\cite{B153}. Following Wilson's ingenious idea of reducing the number of
states by assuming a conduction band with constant density of states we
subdivide the energy band into cells $\mathfrak{C}_{\nu}$ which are
represented by a single state $\widetilde{c}_{\nu}\left(  \mathbf{r}\right)
$. Wilson also normalized the energy and momentum of the electron states.
Since the presented arguments are not based on numerical calculations but on
the discussion of physical properties we use in this paper (and only this
paper) regular electron momenta $k$ and energies $E$. The wave vector $k$ for
the conduction band extends from $0\leq k\leq2k_{F}$. Like Wilson we use a
linear dispersion relation between the energy $E$ and the wave vector $k.$ In
contrast to Wilson we don't count the energy with respect to the Fermi level
but set $E=\hbar v_{F}k$ so that $0\leq E\leq2E_{F}$. The lower (occupied)
half of the energy band is divided at the energies $E_{1}$, $E_{2}$, ...,
$E_{n}$ where $E_{\nu}=$ $E_{F}\left(  1-\frac{1}{2^{\nu}}\right)  ,$ forming
energy cells $\mathfrak{C}_{\nu}=\left(  E_{\nu-1},E_{\nu}\right)  $ or
$\left(  k_{\nu-1},k_{\nu}\right)  .$ The width of the energy cells is
$\left(  E_{\nu}-E_{\nu-1}\right)  =E_{F}2^{-\nu}$ $=\Delta_{\nu}E_{F}$ where
$\Delta_{\nu}=2^{-\nu}$. The energy band above the Fermi level is sub-divided
in a mirrored fashion. Each cell $\mathfrak{C}_{\nu}$ contains $Z_{\nu}$
eigenstates $\widetilde{\varphi}_{j}\left(  \mathbf{r}\right)  $ of the
disordered host. (Throughout this paper we denote the wave function
$\widetilde{\varphi}_{j}\left(  \mathbf{r}\right)  $, the creation operator
$\varphi_{j}^{\dagger}$ and the annihilation operator $\widehat{\varphi}_{j}$
of the same state by the same symbol $\varphi_{j}$ with different decorations).

Following Wilson we represent all the states in a cell $\mathfrak{C}_{\nu}$ by
a linear superposition of the eigenstates $\widetilde{\varphi}_{j}\left(
\mathbf{r}\right)  $
\[
\widetilde{c}_{\nu}\left(  \mathbf{r}\right)  =\frac{1}{A}%
{\displaystyle\sum_{j\in\mathfrak{C}_{\nu}}}
\widetilde{\varphi}_{j}^{\ast}\left(  \mathbf{0}\right)  \widetilde{\varphi
}_{j}\left(  \mathbf{r}\right)
\begin{tabular}
[c]{l}%
,\
\end{tabular}
\ \ \ \ \ \ A^{2}=%
{\displaystyle\sum_{j\in\mathfrak{C}_{\nu}}}
\left\vert \widetilde{\varphi}_{j}\left(  \mathbf{0}\right)  \right\vert ^{2}%
\]
where the summation goes over all states $j$ with energy $\varepsilon_{j}$ in
the cell $\mathfrak{C}_{\nu}.$\ Then the amplitude of $\widetilde{c}_{\nu
}\left(  \mathbf{r}\right)  $ at the origin is
\[
\widetilde{c}_{\nu}\left(  \mathbf{0}\right)  =\sqrt{%
{\displaystyle\sum_{j\in\mathfrak{C}_{n}}}
\left\vert \widetilde{\varphi}_{j}\left(  \mathbf{0}\right)  \right\vert ^{2}}%
\]
Since the cell $\mathfrak{C}_{\nu}$ originally contained $Z_{\nu}$ electron
\ states there are $\left(  Z_{\nu}-1\right)  $ states left. These states can
be orthonormalized with respect to each other and the state $\widetilde
{c}_{\nu}$. The resulting states we might call $\widetilde{c}_{\nu,\mu}$. If
the composition of such a state $\widetilde{c}_{\nu,\mu}\left(  \mathbf{r}%
\right)  $ is
\[
\widetilde{c}_{\nu,\mu}\left(  \mathbf{r}\right)  =%
{\displaystyle\sum_{j\in\mathfrak{C}_{\nu}}}
\alpha_{\nu,\mu}^{j}\widetilde{\varphi}_{j}\left(  \mathbf{r}\right)
\]
then the orthogonality condition
\[
0=\left\langle \widetilde{c}_{\nu}|\widetilde{c}_{\nu,\mu}\right\rangle =\int
d^{3}\mathbf{r}\frac{1}{A}%
{\displaystyle\sum_{j\in\mathfrak{C}_{\nu}}}
\widetilde{\varphi}_{j}\left(  \mathbf{0}\right)  \widetilde{\varphi}%
_{j}^{\ast}\left(  \mathbf{r}\right)
{\displaystyle\sum_{j^{\prime}\in\mathfrak{C}_{\nu}}}
\alpha_{\nu,\mu}^{j^{\prime}}\widetilde{\varphi}_{j^{\prime}}\left(
\mathbf{r}\right)  =\frac{1}{A}%
{\displaystyle\sum_{j\in\mathfrak{C}_{\nu}}}
\alpha_{\nu,\mu}^{j}\widetilde{\varphi}_{j}\left(  \mathbf{0}\right)
=\frac{1}{A}\widetilde{c}_{\nu,\mu}\left(  \mathbf{0}\right)
\]
yields that all the other states $\widetilde{c}_{\nu,\mu}\left(
\mathbf{r}\right)  $ have vanishing amplitude at the origin and don't interact
with the magnetic impurity.

Let the volume of the host be $V$ and the total number of electrons be $2Z$
(i.e. $Z$ per spin). Then the average electron density per spin is $n_{0}%
=Z/V$. This electron density is in good approximation homogeneously
distributed over the energy range from the bottom of the band up to the Fermi
energy, i.e. between $0$ and $E_{F}$ in the energy band. Since the width of
the energy cell $\mathfrak{C}_{\nu}$ is $\Delta_{\nu}E_{F}$ then this energy
cell contributes the fraction $\Delta_{\nu}n_{0}$ to the electron density. On
the other hand the density of the Wilson state $\widetilde{c}_{\nu}\left(
\mathbf{r}\right)  $ at the origin is $%
{\displaystyle\sum_{j}}
\left\vert \widetilde{\varphi}_{j}\left(  \mathbf{0}\right)  \right\vert ^{2}%
$. Since all the remaining band states in the cell $\mathfrak{C}_{\nu}$ have
zero density at the origin it follows that
\[%
{\displaystyle\sum_{j\in\mathfrak{C}_{n}}}
\left\vert \widetilde{\varphi}_{j}\left(  \mathbf{0}\right)  \right\vert
^{2}=\Delta_{\nu}n_{0}=A^{2}%
\]

Then the amplitude of the Wilson state at the impurity is $A=\sqrt{\Delta
_{\nu}n_{0}}$. This is the same amplitude as in the pure host. As long as we
have a large volume (so that the energy level spacing of the $\widetilde
{\varphi}_{j}\left(  \mathbf{r}\right)  $ is much smaller than the Kondo
energy) and neglect the small fluctuations in the local electron density due
to the disorder we arrive at essentially the same Hamiltonian (equ.
(\ref{hfa0})) for the Wilson states as in the pure host. This is because the
conduction electrons enter the Hamiltonian only through the energy of the
Wilson states and their amplitude $\widetilde{c}_{\nu}\left(  \mathbf{0}%
\right)  $ at the impurity. The Hamiltonian of the FA-impurity is given by%
\begin{equation}
H_{FA}=%
{\textstyle\sum_{\sigma}}
\left\{  \sum_{\nu=1}^{N}\varepsilon_{\nu}c_{\nu\sigma}^{\dag}c_{\nu\sigma
}+E_{d}d_{\sigma}^{\dag}d_{\sigma}+\sum_{\nu=1}^{N}V_{sd}(\nu)[d_{\sigma
}^{\dag}c_{\nu\sigma}+c_{\nu\sigma}^{\dag}d_{\sigma}]\right\}  +Un_{d,\uparrow
}n_{d,\downarrow} \label{hfa0}%
\end{equation}
where $c_{\nu\sigma}^{\dagger}$ and $d_{\sigma}^{\dagger}$ are the creation
operators for the Wilson and d-states. $V_{sd}\left(  \nu\right)  $ is the
s-d-coupling which is proportional to $\widetilde{c}_{\nu\sigma}\left(
\mathbf{0}\right)  $ and $U$ is the exchange interaction.

Therefore the mathematical form of the Kondo ground state is (in this
approximation) the same for the disordered host as for the pure one. But one
has to keep in mind that the almost identical properties of the Wilson states
for the disordered and pure host are restricted to $\mathbf{r=0}$. At larger
distances $r>>l$ from the magnetic impurity the wave functions of the Wilson
states for the pure and disordered host are quite different. This has
important consequences for the extent of the Kondo cloud.

However, in the first step of the calculation we can essentially take the
Wilson Hamiltonian of the FA-impurity in the Anderson model for the pure host.
Since this Hamiltonian is already solved we can use this solution for the
future discussion. We present the solution in the FAIR description
\cite{B151}.
\begin{align}
\Psi_{SS}  &  =\left[  Aa_{0,\uparrow}^{\dagger}b_{0,\downarrow}^{\dagger
}+Ba_{0,\uparrow}^{\dagger}d_{\downarrow}^{\dagger}+Cd_{\uparrow}^{\dagger
}b_{0,\downarrow}^{\dagger}+Dd_{\uparrow}^{\dagger}d_{\downarrow}^{\dagger
}\right]  \left\vert \mathbf{0}_{a\uparrow}\mathbf{0}_{b\downarrow
}\right\rangle \label{Psi_SS}\\
&  +\left[  Ab_{0,\uparrow}^{\dagger}a_{0,\downarrow}^{\dagger}+Cb_{0,\uparrow
}^{\dagger}d_{\downarrow}^{\dagger}+Bd_{\uparrow}^{\dagger}a_{0,\downarrow
}^{\dagger}+Dd_{\uparrow}^{\dagger}d_{\downarrow}^{\dagger}\right]  \left\vert
\mathbf{0}_{b\uparrow}\mathbf{0}_{a\downarrow}\right\rangle \nonumber
\end{align}

where%
\[
\left\vert \mathbf{0}_{a\uparrow}\mathbf{0}_{b\downarrow}\right\rangle
=\prod_{i=1}^{n-1}a_{i,\uparrow}^{\dagger}\prod_{i=1}^{n-1}b_{i,\downarrow
}^{\dagger}\left\vert \Phi_{0}\right\rangle
\begin{tabular}
[c]{l}%
, \
\end{tabular}
\ \ \ \ \ \ \ \ \left\vert \mathbf{0}_{b\uparrow}\mathbf{0}_{a\downarrow
}\right\rangle =\prod_{i=1}^{n-1}b_{i,\uparrow}^{\dagger}\prod_{i=1}%
^{n-1}a_{i,\downarrow}^{\dagger}\left\vert \Phi_{0}\right\rangle
\]
are the half-filled polarized conduction bands, $\left\vert \Phi
_{0}\right\rangle $ is the vacuum state \footnote{We arrive at the final
$\Psi_{SS}$ by initially building the two \emph{fair} states $a_{0}^{\dagger}$
and $b_{0}^{\dagger}$ out of the Wilson basis $\left\{  c_{\nu}^{\dagger
}\right\}  $. The \emph{fair} states define the full bases $\left\{
a_{i}^{\dagger}\right\}  $ and $\left\{  b_{i}^{\dagger}\right\}  $ uniquely.
Then the energy expectation value of $\Psi_{SS}$ is calculated and the
\emph{fair} states $a_{0}^{\dagger}$ and $b_{0}^{\dagger}$ are varied until
$\Psi_{SS}$ with the lowest energy is obtained. The initial $a_{0}^{\dagger}$
and $b_{0}^{\dagger}$ can be arbitrary but different. An educated guess
reduces the variation time. Details are in ref. \cite{B187}.}.

For the calculation of the Kondo cloud we divide the Kondo ground state in
equ. (\ref{Psi_SS}) in two magnetic components, the top and the bottom part of
the singlet solution. In zero magnetic field they have opposite s-electron
polarization. This polarization is distributed over the occupied states which
are composed of Wilson states.

In the disordered case $\widetilde{c}_{\nu}\left(  \mathbf{r}\right)
,\widetilde{a}_{i}\left(  \mathbf{r}\right)  $ and $\widetilde{b}_{i}\left(
\mathbf{r}\right)  $ are composed of the eigenstates of the disordered host
which are in general unknown. Therefore the tricky part in the disordered host
is to gain the essential information about the wave function of $\widetilde
{c}_{\nu}\left(  \mathbf{r}\right)  $.

\section{The Pure Case}

In the following we want to compare the pure host with the dirty host. This
consideration is more transparent when we don't use Wilson's dimensionless
momenta and energies but the standard variables $k$ and $E$ for the momentum
and energy of the electron.

In a sphere of radius $R$ the normalized eigenstates with finite amplitude at
the origin have the form%
\begin{equation}
\widetilde{\varphi}_{j}\left(  r\right)  =\frac{1}{\sqrt{2\pi R}}\frac{1}%
{r}\sin k_{j}r \label{fi_j}%
\end{equation}
where $k_{j}=j\pi/R$ is a standard wave number and $k_{\Delta}=\pi/R$ is the
step width of $k_{j}$. The averaged density $\rho_{0}\left(  r\right)  $
(averaged over $\sin^{2}\left(  k_{j}r\right)  $), integrated over a spherical
shell of thickness $dr$ for $r>>2\pi/k_{j}$ is
\[
4\pi r^{2}\rho_{0}\left(  r\right)  dr=4\pi r^{2}dr\frac{1}{2\pi R}\frac
{1}{r^{2}}\frac{1}{2}=\frac{dr}{R}%
\]

The state $\widetilde{\varphi}_{j}\left(  r\right)  $ can be split into an
incoming and an outgoing wave. The latter has the form
\[
\widetilde{\varphi}_{j,o}\left(  r\right)  =\frac{1}{\sqrt{2\pi R}}\frac
{1}{2i}\frac{1}{r}\exp\left(  ik_{j}r\right)
\]
and the incoming part is the conjugate complex state. The outflow of the
outgoing wave is
\[
J=4\pi r^{2}\left\vert \widetilde{\varphi}_{j,o}\left(  r\right)  \right\vert
^{2}v_{F}=4\pi r^{2}\frac{1}{2\pi R}\frac{1}{4}\frac{1}{r^{2}}v_{F}%
=\frac{v_{F}}{2R}%
\]
The incoming part has the inflow which is equal and opposite. So the density
is equal to $\rho_{0}\left(  r\right)  =2J/v_{F}=1/\left(  4\pi r^{2}R\right)
$

A Wilson state has the form%

\[
\widetilde{c}_{\nu}\left(  r\right)  =\frac{1}{\sqrt{Z_{\nu}}}%
{\displaystyle\sum_{j\in\mathfrak{C}_{\nu}}}
\frac{1}{\sqrt{2\pi R}}\frac{1}{r}\sin k_{j}r=>
\]
with $Z_{\nu}=k_{F}R2^{-\nu}/\pi=k_{F}R\Delta_{\nu}/\pi$ and $%
{\displaystyle\sum_{j}}
$=%
$>$%
$\int dk/k_{\Delta}$=$\frac{R}{\pi}\int dk$%
\[
=>\frac{1}{\pi}\sqrt{\frac{1}{2k_{F}\Delta_{\nu}}}\int_{k_{F}\left(
1-2^{\nu-1}\right)  }^{k_{F}\left(  1-2^{-\nu}\right)  }\frac{1}{r}\sin\left(
kr\right)  dk
\]
Integration and some manipulations yield%
\[
\widetilde{c}_{\nu}\left(  r\right)  =\frac{1}{\pi}\sqrt{k_{F}\frac
{\Delta_{\nu}}{2}}\frac{1}{r}\sin\left[  k_{F}\left(  1-\frac{2^{-\nu}%
+2^{-\nu-1}}{2}\right)  r\right]  \left\{  \frac{1}{k_{F}\frac{\Delta_{\nu}%
}{2}r}\sin\left[  k_{F}\frac{\Delta_{\nu}}{2}r\right]  \right\}
\]
The main contribution of this wave function is in the range $0\leq
r\leq2/\left(  \Delta_{\nu}k_{F}\right)  $ $\thickapprox2^{\nu+1}/k_{F}$ where
the term in the right wavy bracket is of the order of one. In this range the
term in the left square bracket is roughly normalized and the density
$\rho_{\nu}^{0}\left(  r\right)  $ for $r>>2/\left(  \Delta_{\nu}k_{F}\right)
$ approaches zero with increasing $r$ because of the $r^{-2}$ radial
dependence of $\widetilde{c}_{\nu}\left(  r\right)  $.

\subsection{Calculation of the polarization}

The Kondo ground state is an entanglement of two magnetic components with
opposite net moment of the d-state. The Kondo cloud is given by the
polarization of one of the two magnetic components. (We construct the ground
state so that the first half in equ. (\ref{Psi_SS}) has a net negative
d-spin.
\begin{equation}
\Psi_{\downarrow}=\left[  Aa_{0,\uparrow}^{\dagger}b_{0,\downarrow}^{\dagger
}+Ba_{0,\uparrow}^{\dagger}d_{\downarrow}^{\dagger}+Cd_{\uparrow}^{\dagger
}b_{0,\downarrow}^{\dagger}+Dd_{\uparrow}^{\dagger}d_{\downarrow}^{\dagger
}\right]  \left\vert \mathbf{0}_{a\uparrow}\mathbf{0}_{b\downarrow
}\right\rangle \label{Psi_MS}%
\end{equation}
Here $\left\vert \mathbf{0}_{a\uparrow}\mathbf{0}_{b\downarrow}\right\rangle
=\left\vert \mathbf{0}_{a\uparrow}\right\rangle \left\vert \mathbf{0}%
_{b\downarrow}\right\rangle $ with $\left\vert \mathbf{0}_{a\uparrow
}\right\rangle =%
{\displaystyle\prod\limits_{i=1}^{n}}
a_{i\uparrow}^{\dagger}\left\vert \Phi_{0}\right\rangle $ is the
(anti-symmetric) product of the occupied wave functions $\widetilde
{a}_{i\uparrow}\left(  \mathbf{r}\right)  $ which are composed of Wilson
states%
\begin{equation}
\widetilde{a}_{i}\left(  \mathbf{r}\right)  =%
{\displaystyle\sum_{\nu}}
K_{i}^{\nu}\widetilde{c}_{\nu\uparrow}\left(  \mathbf{r}\right)
\label{Fair_bd}%
\end{equation}
The matrix $K_{i}^{\nu}$ has been determined in the process of deriving the
FAIR solution (\ref{Psi_SS}).
\begin{equation}
\widetilde{a}_{i}\left(  r\right)  =\frac{1}{\pi}\frac{1}{r^{2}}%
{\displaystyle\sum_{\nu}}
\sqrt{\frac{2}{\Delta_{\nu}k_{F}}}K_{i}^{\nu}\sin\left[  k_{F}\left(
1-\frac{2^{-\nu}+2^{-\nu-1}}{2}\right)  r\right]  \sin\left[  k_{F}%
\frac{\Delta_{\nu}}{2}r\right]
\end{equation}

The orbital part of $\widetilde{a}_{i\uparrow}\left(  \mathbf{r}\right)  $ and
$\widetilde{a}_{i\downarrow}\left(  \mathbf{r}\right)  $ are identical. Then
the contribution of any wave fuction $\widetilde{a}_{i\uparrow}\left(
\mathbf{r}\right)  $ in the occupied FAIR band $\left\vert \mathbf{0}%
_{a\uparrow}\right\rangle $ to the spin polarization of $\Psi_{\downarrow}$ is
$\frac{1}{2}\left\vert \widetilde{a}_{i}\left(  r\right)  \right\vert ^{2}$.
The definitions of the wave functions $\widetilde{b}_{i}\left(  \mathbf{r}%
\right)  $ and their contribution to the polarization are equivalent.

The total polarization of the s-electrons in $\Psi_{\downarrow}$ is
\begin{align*}
p^{0}\left(  r\right)  dr  &  =\frac{1}{2}\left[  \left\vert A\right\vert
^{2}\left(  \left\vert \widetilde{a}_{0}\left(  \mathbf{r}\right)  \right\vert
^{2}-\left\vert \widetilde{b}_{0}\left(  \mathbf{r}\right)  \right\vert
^{2}\right)  +\left\vert B\right\vert ^{2}\left\vert \widetilde{a}_{0}\left(
\mathbf{r}\right)  \right\vert ^{2}-\left\vert C\right\vert ^{2}\left\vert
\widetilde{b}_{0}\left(  \mathbf{r}\right)  \right\vert ^{2}\right]  dr\\
&  +\frac{1}{2}%
{\displaystyle\sum_{i=1}^{n}}
\left[  \left\vert \widetilde{a}_{i}\left(  \mathbf{r}\right)  \right\vert
^{2}-\left\vert \widetilde{b}_{i}\left(  \mathbf{r}\right)  \right\vert
^{2}\right]  dr
\end{align*}

Within the FAIR theory this polarization $p^{0}\left(  r\right)  $ in the pure
host has been calculated \cite{B177}.

\section{Disordered Host}

We consider a large weakly disordered host with $k_{F}l>>1$. Furthermore the
magnetic impurity (which we locate at the origin) should not be close to a
normal impurity. In that case in the range $r<l/2$ the Wilson states
$\widetilde{c}_{\nu}\left(  \mathbf{r}\right)  $ possess essentially the same
wave functions as in the pure case. Within the cage of the nearest impurities
the electrons are free and can be expressed as superpositions of plane waves.
Let us assume that very close to the magnetic impurity the eigenstate of the
disordered host $\widetilde{c}_{j}\left(  \mathbf{r}\right)  $ has the
asymptotic form $\widetilde{\varphi}_{\mathbf{k}}\left(  \mathbf{r}\right)
\cong\alpha_{\mathbf{k}}e^{i\mathbf{k}\left(  \mathbf{r-r}_{0}\right)  }$.
Then it contributes to the (appropriate) Wilson state $\widetilde{c}_{\nu
}\left(  \mathbf{r}\right)  $ the component $\widetilde{\varphi}_{\mathbf{k}%
}\left(  \mathbf{r}\right)  \widetilde{\varphi}_{\mathbf{k}}^{\ast}\left(
\mathbf{0}\right)  \cong\left\vert \alpha_{\mathbf{k}}\right\vert
^{2}e^{i\mathbf{kr}}$

The plane wave $\alpha_{\mathbf{k}}e^{i\mathbf{kr}}$ consists of Bessel
functions
\[
\left\vert \alpha_{\mathbf{k}}\right\vert ^{2}e^{i\mathbf{kr}}=\left\vert
\alpha_{\mathbf{k}}\right\vert ^{2}4\pi%
{\displaystyle\sum_{l=0}^{\infty}}
{\displaystyle\sum_{m=-l}^{l}}
i^{l}j_{l}\left(  kr\right)  Y_{l}^{m\ast}\left(  \theta_{\mathbf{k}%
}\mathbf{,\phi}_{\mathbf{k}}\right)  Y_{l}^{m}\left(  \theta_{\mathbf{r}}%
,\phi_{\mathbf{r}}\right)
\]
Only the $l=0$ component is non-zero at the origin and equal to $\left\vert
\alpha_{\mathbf{k}}\right\vert ^{2}j_{0}\left(  kr\right)  =\left\vert
\alpha_{\mathbf{k}}\right\vert ^{2}\sin\left(  kr\right)  /\left(  kr\right)
$. Summing over all eigenstates in the energy cell $\mathfrak{C}_{\nu}$ yields
a constructive interference for the $l=0$ components with a total amplitude of
$%
{\displaystyle\sum_{\mathbf{k}\ni\mathfrak{C}_{\nu}}}
\left\vert \alpha_{\mathbf{k}}\right\vert ^{2}\sin\left(  kr\right)  /\left(
kr\right)  $ where $%
{\displaystyle\sum_{\mathbf{k}\in\mathfrak{C}_{\nu}}}
\left\vert \alpha_{\mathbf{k}}\right\vert ^{2}\cong\sqrt{\Delta_{\nu}n_{0}}$.
The contributions of the $l>0$ angular momenta cancel to zero in first
approximation. This argument also works if the asymptotic form of
$\widetilde{\varphi}_{\mathbf{k}}\left(  \mathbf{r}\right)  $ is a standing wave.

When we sum over all disordered states in the energy cell $\mathfrak{C}_{\nu}$
we obtain within $r<l$ essentially the same state as in the pure host. It will
not be perfectly spherical and is slightly disturbed by the back scattering
from the nearest impurities. If one would perform impurity averaging at this
point the state would become spherically symmetric, but it would also decay
exponentially with increasing distance. However, we will perform the impurity
averaging in a later stage.

First we recall that $\widetilde{c}_{\nu}\left(  \mathbf{r}\right)  $ (in the
pure host) represents a superposition of a spherical incoming and outgoing
wave. In the pure case the trajectories of the waves are radial beams. In the
disordered host the scattering folds these trajectories. At each impurity the
trajectory is split and folded (see Fig.1) and a phase shift is attached to
the wave function along the trajectory.
\begin{align*}
&
{\includegraphics[
height=3.4097in,
width=2.9082in
]%
{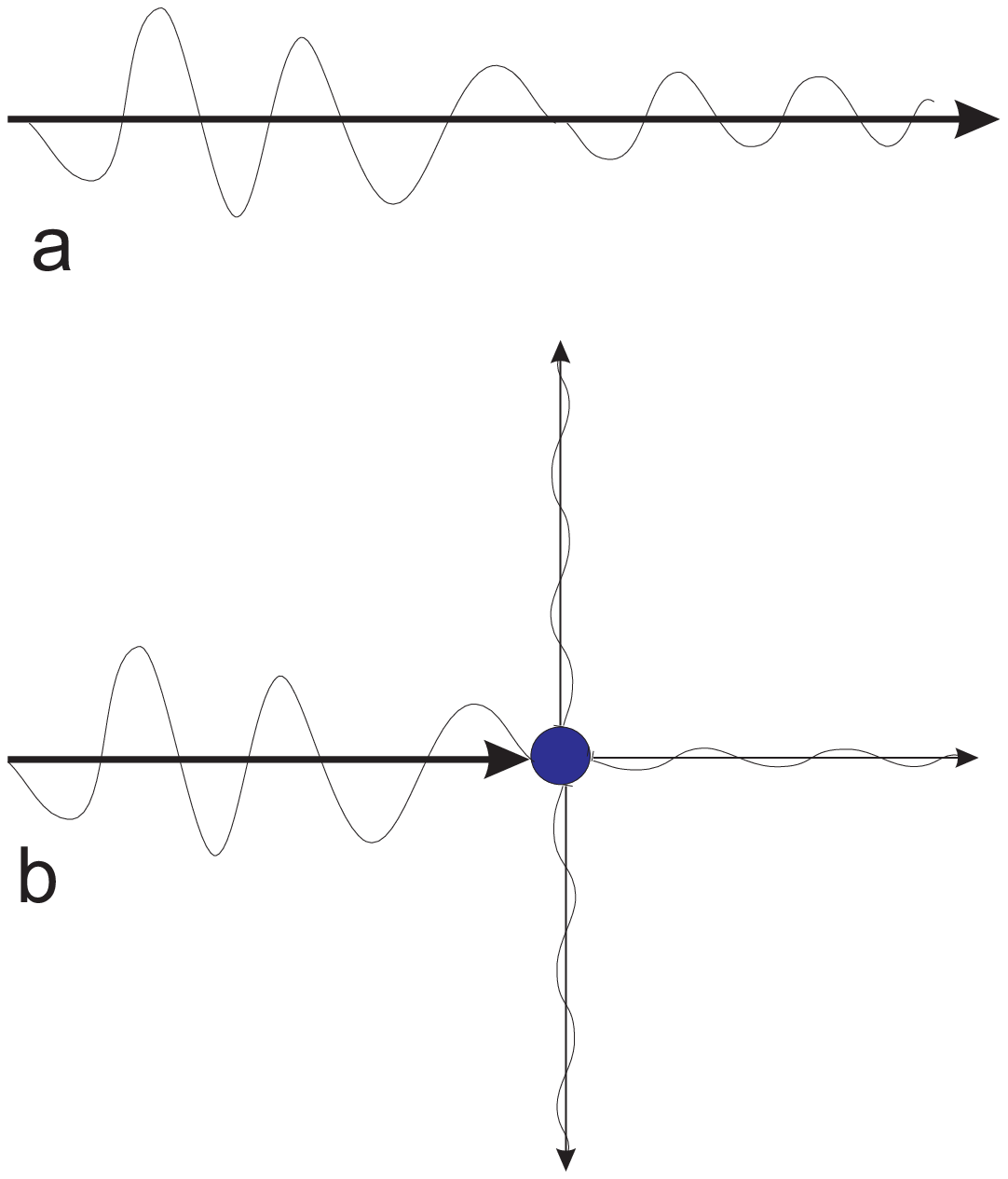}%
}%
\\
&
\begin{tabular}
[c]{l}%
Fig.1: a) The radial trajectory in a pure host.\\
b) In the disordered host the trajectory splits\\
at each impurity into different directions and\\
picks up a phase shift. In both cases the wave\\
function is plotted along the trajectories.
\end{tabular}
\end{align*}

Before we discuss the density distribution we proceed to the basis states of
the FAIR band, for example $\widetilde{a}_{i}\left(  \mathbf{r}\right)  $. In
the pure case we have calculated numerically the wave function $\widetilde
{a}_{i}\left(  r\right)  $ using the Wilson states. Its value along the radial
trajectory is known. In the disordered case we use the same same wave function
$\widetilde{a}_{i}\left(  r_{t}\right)  $ along the trajectory where $r_{t}$
is the path along the trajectory (not the distance from the impurity). Again
at each impurity the trajectory (and the wave function) is split, its
direction is changed and a phase shift is attached to the wave function along
the next leg of the trajectory.

In Fig.2 a few folded trajectories are shown. While the (properly) integrated
density of $\left\vert \widetilde{a}_{i}\left(  r_{t}\right)  \right\vert
^{2}$ over each leg of all trajectories is still normalized the crossing
between different trajectories yields interferences.%

\begin{align*}
&
{\includegraphics[
height=3.3391in,
width=3.3391in
]%
{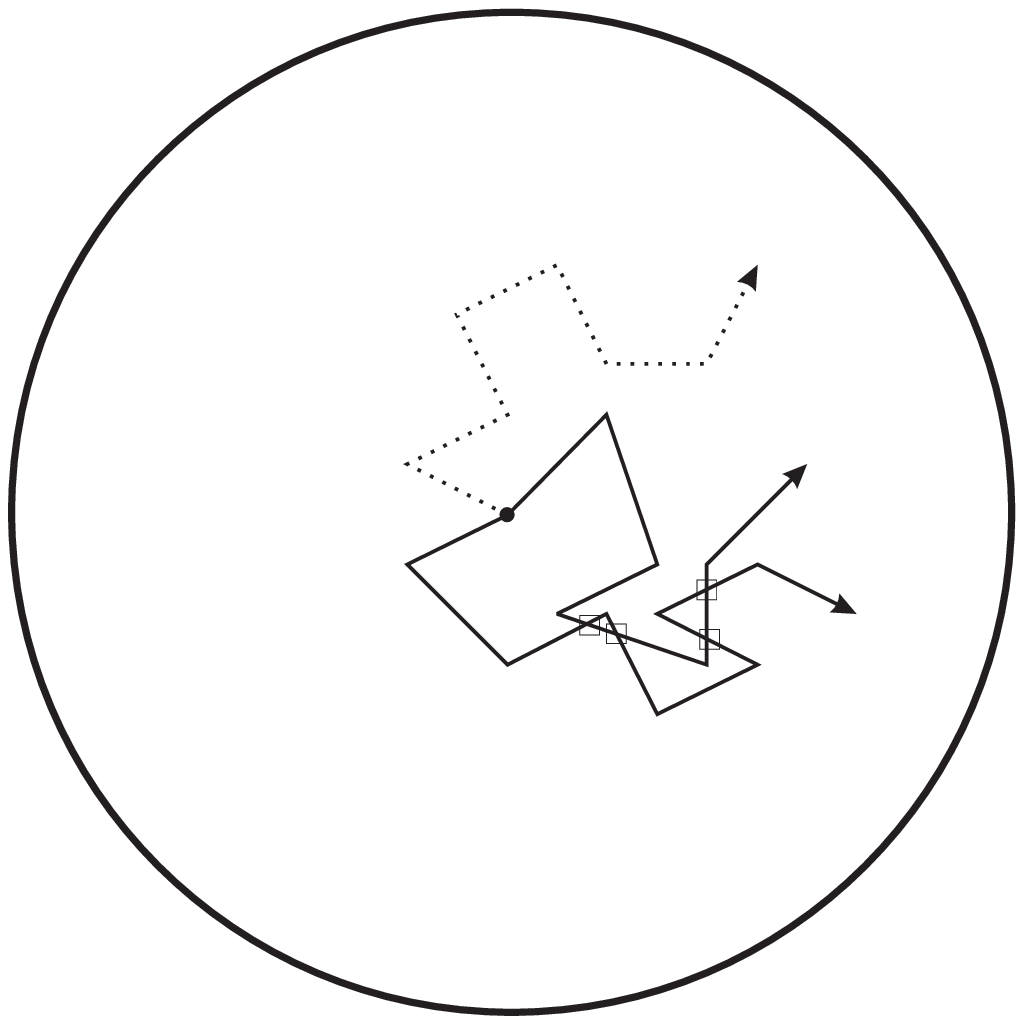}%
}%
\\
&
\begin{tabular}
[c]{l}%
Fig.2: Random propagation in a disordered host\\
with overlapping trajectories.
\end{tabular}
\end{align*}

This random propagation represents a quantum diffusion. A trajectory which in
the pure host reaches the distance $r_{0}$ from the origin will have in an
average $r_{0}/l$ collisions and reaches only a distance of $\sqrt{r_{0}%
/l}l=\sqrt{r_{0}l}$ from the origin. The detailed distribution of the charge
$\left\vert \widetilde{a}_{i}\left(  \mathbf{r}\right)  \right\vert ^{2}$ in
the disordered host depends on the distribution of the impurities. Therefore
we perform an impurity averaging. This averages the interferences to zero. Now
we can calculate the averaged charge distribution of our FAIR state
$\widetilde{a}_{i}\left(  r\right)  $.

We consider in the pure host the in and outgoing waves of $\widetilde{a}%
_{i}\left(  r\right)  $ in the spherical shell between $r_{0}$ and
$r_{0}+dr_{0}$. Let the density at the distance $r_{0}$ be $\rho_{i}%
^{0}\left(  r_{0}\right)  ,$ then the total charge in the spherical shell of
thickness $dr_{0}$ is $4\pi r_{0}^{2}\rho_{i}^{0}\left(  r_{0}\right)  dr_{0}%
$. In the disordered host this charge will obey a diffusion profile with the
diffusion constant $D=\frac{1}{3}v_{F}l$ yielding
\begin{equation}
d\overline{\rho_{i}^{do}}\left(  r\right)  =4\pi r_{0}^{2}\rho_{i}^{0}\left(
r_{0}\right)  \frac{1}{\left(  \pi Dt_{0}\right)  ^{3/2}}\exp\left(
-\frac{r^{2}}{4Dt_{0}}\right)  dr_{0} \label{ro_do}%
\end{equation}
where $t_{0}$ stands for $r_{0}/v_{F}$ which could be interpreted as a
traveling time. We have to add the contribution from all spherical shells
(which corresponds to an integration over $dr_{0}$). This yields the charge
density $\overline{\rho_{i}^{do}}\left(  r\right)  $ of $\widetilde{a}%
_{i}\left(  r\right)  $ in the disordered host in terms of the charge density
in the pure host $\rho_{i}^{0}\left(  r_{0}\right)  $.

The density for all FAIR states $\widetilde{a}_{i}\left(  r\right)  $, and
$\widetilde{b}_{i}\left(  r\right)  $ has been calculated in the pure case in
\cite{B177}. In complete analogy these densities in the pure host can be
translated into the disordered case using relation (\ref{ro_do}). The total
averaged polarization $\overline{p_{do}}\left(  r\right)  $ of the state
$\Psi_{\downarrow}$ is then
\begin{equation}
\overline{p_{do}}\left(  r\right)  =4\pi\int_{0}^{\infty}r_{0}^{2}p_{0}\left(
r_{0}\right)  \frac{1}{\left(  \pi lr_{0}/3\right)  ^{3/2}}\exp\left(
-\frac{3r^{2}}{4lr_{0}}\right)  dr_{0} \label{SE_a}%
\end{equation}

Within this model the Kondo cloud polarization in a disordered host can be
obtained from the polarization in a pure host by means of equ. (\ref{SE_a}).
Any polarization $p_{0}\left(  r\right)  $ in the pure host can be transferred
into the disordered host.

The range of the polarization in a disordered host is of the order of
$\sqrt{\xi_{K}l}$ where $\xi_{K}$ is the Kondo length in the pure host.

For a Kondo system with a Kondo temperature of $T_{K}\thickapprox2K$ and a
host with a Fermi velocity of about $v_{F}\thickapprox3\times10^{6}m/s$ the
Kondo length (in a pure host) is about $\xi_{K}=10\mu$. This incloses a sphere
of volume $4\times10^{3}\mu^{3}$. If the host has a mean free path of $l=10nm$
then one obtains for the dirty host a Kondo length of about $\allowbreak
\xi_{K}^{do}\thickapprox0.3\mu$. Now the Kondo cloud is distributed over
volume which is smaller by a factor of $4\times10^{4}$. This improves the
chance to detect the Kondo cloud experimentally although some of the previous
experimental methods might not be applicable in the disordered host.

\section{Conclusions}

The Kondo cloud in a weakly disordered host is investigated. The two magnetic
components which are entangled in the ground state are artificially separated
and the spin polarization of one of the components is calculated. If the
magnetic impurity is well separated from the non-magnetic impurities then
within a sphere of radius $l/2$ (i.e. half the mean free path) the
ground-state wave functions in the disordered host are very close to those of
the pure host. These wave functions are superpositions of incoming and
outgoing spherical waves. In the pure host their propagation is radial and the
difference between their densities for spin up and down give the polarization
of the Kondo cloud. In the disordered host the trajectories are split at the
impurities and represent quantum diffusion. Averaging over the impurity
position permits us to calculate the polarization in the disordered host from
the known polarization in the pure host. The spatial extension is reduced from
the Kondo length $\xi_{K}$ in the pure host to $\sqrt{\xi_{K}l}$ in the
disordered host.

\end{document}